\documentclass[aps,pra,twocolumn,amsmath,amssymb,showpacs]{revtex4}

\newcommand{\ket}[1]{|{#1}\rangle}
\newcommand{\bra}[1]{\langle{#1}|}
\usepackage[dvips]{graphicx}

\begin{document}

\title{Frequency and temporal effects in linear optical quantum computing}

\author{Peter P. Rohde}
\email[]{rohde@physics.uq.edu.au}
\homepage{http://www.physics.uq.edu.au/people/rohde/}
\author{Timothy C. Ralph}
\affiliation{Centre for Quantum Computer Technology, Department of Physics\\University of Queensland, QLD 4072, Australia}

\date{\today}

\begin{abstract}
Typically linear optical quantum computing (LOQC) models assume that all input photons are completely indistinguishable. In practice there will inevitably be non-idealities associated with the photons and the experimental setup which will introduce a degree of distinguishability between photons. We consider a non-deterministic optical controlled-NOT gate, a fundamental LOQC gate, and examine the effect of temporal and spectral distinguishability on its operation. We also consider the effect of utilizing non-ideal photon counters, which have finite bandwidth and time response.
\end{abstract}

\pacs{03.67.Lx,42.50.-p}

\maketitle

\section{INTRODUCTION}
It has been shown that efficient quantum computation can be implemented using only linear optical elements, single photon sources and photon counters \cite{bib:KLM01}. This scheme is referred to as linear optical quantum computing (LOQC) and has recently been subject to several in-principle experimental demonstrations \cite{bib:Pittman03,bib:OBrien03,bib:Zhao05,bib:Gasparoni04}.

In the LOQC model qubits are encoded as the presence in one of two modes of a single photon (dual-rail logic). Thus a logical zero can be written  $\ket{0}_L\equiv\ket{01}$ where $\ket{ij}$ is a two mode state in which $i$ is the occupation number of the first mode and and $j$ is the occupation number of the second mode. Similarly a logical one is written  $\ket{1}_L\equiv\ket{10}$. Often these modes are taken to be the vertical and horizontal polarization modes of a single spatial mode. The non-linearities which are necessary for quantum computation are introduced via a conditioning or post-selection process whereby the outputs from the system are only accepted if certain states are detected at extra ancillary outputs. This post-selection process results in non-deterministic quantum gate operation (\mbox{\emph{i.e.}} gates have a certain probability of failing), however, through the use of a teleportation protocol, it is possible in-principle to boost the probability of success arbitrarily close to unity \cite{bib:KLM01}.

It is important to understand how critically non-ideal factors affect basic gate operation. Previous studies have looked at the effects of beamsplitter parameters \cite{bib:Ralph02}, detection \cite{bib:Glancy02} and ancilla \cite{bib:Lund03} efficiencies on various gate designs.

A key ingredient of the post-selection process is the presence of non-classical interference due to photon indistinguishability. Thus one would expect gate operation to be compromised if the input photons are distinguishable in some way. In this paper we consider such a problem by analyzing the effect of non-identical frequency distributions and unsynchronized arrival times on the operation of a non-deterministic 2-qubit LOQC gate. Some previous work has been done on the effect of unsynchronized arrival times in the context of quantum dot single photon sources \cite{bib:Kiraz04}. We begin by introducing a single photon time/frequency representation. We then illustrate our techniques and some basic physical effects by the example of 2 photon interference at a 50/50 beamsplitter. In section \ref{sec:CNOT_gate} we reapply these techniques to the more complicated case of the 2-qubit gate and also give consideration to frequency and temporal effects within the photon-counters. In section \ref{sec:conclusion} we conclude.

\section{REPRESENTATION OF SINGLE PHOTON STATES}
Typically when we employ single photon input states in the LOQC model we implicitly assume that all photons input into the system are completely indistinguishable. Hence, although all of the photons necessarily exhibit distributions in frequency space, it suffices to treat them all identically and express them in Dirac notation simply as $\ket{1}$, which ignores the exact form of their distribution. If, however, the input states are not completely indistinguishable then such a notation is insufficient and we must revert to a more general representation for single photon states which allows for their individual distributions. This can be modeled by adopting the notation
\begin{equation} \label{eq:distributed_state}
\ket{1} \to \int \alpha _\omega \ket{1}_\omega \, \mathrm{d}\omega = \Big( \int \alpha _\omega \tilde{\textbf{a}}_ \omega ^\dag \, \mathrm{d}\omega \Big) \ket{0}
\end{equation}
where $\alpha_\omega$ is a normalized function describing the distribution of the photon in frequency space. $\ket{1}_\omega$ is the single frequency, single-photon state. Similarly, $\tilde{\textbf{a}}_\omega ^\dag$ is the single frequency creation operator. The integral is over all frequency space. In choosing this representation of frequency distributed states, we have implicitly assumed that inputs are pure states. Consequently pulses are transform limited and the corresponding time-domain representation of any distribution can be determined by taking the inverse Fourier transform of the distribution function. We will discuss mixed states at the end of this section.

By choosing different forms for $\alpha_\omega$, photons produced by various physical systems can be modeled. We now consider some examples. The output from an optical cavity is characterized by a Lorentzian distribution, which has the form
\begin{equation} \label{eq:lorentzian_distribution}
\alpha _\omega = \sqrt {\frac{\kappa}{\pi}}\frac{1}{\kappa+i\omega}
\end{equation}
where $\kappa$ is a parameter related to the bandwidth of the distribution. Thus we would expect that a single photon source, comprising a single photon emitter in an optical cavity, would be approximately characterized by a Lorentzian distribution. We say approximately because we assume that there is no uncertainty in the time of emission of the photon by the photon emitter. In reality there would be some uncertainty associated with this and the distribution of the photon emitted from the cavity would be represented as a convolution of this distribution and the cavity's response. However, assuming that the time-frame in which photons are placed into the cavity by the photon emitter is much smaller than the decay time of the cavity, a Lorentzian distribution is a valid approximation of the response. Examples are quantum dot \cite{bib:Santori02} and cavity QED \cite{bib:Legero03} based photon sources. The Lorentzian distribution has the property that, although two-sided in the frequency-domain, it is one-sided in the time-domain.

Another possible single photon source is the non-degenerate parametric down converter, frequently used in LOQC experiments. This is a device which takes a single photon as an input and produces two photons, each with half the frequency of the original photon, but in orthogonal spatial or polarization modes, at its outputs. Considerable work has been done on the frequency distributions of down-conversion sources \cite{bib:Rubin94,bib:Walmsley01}. The form of the output from a down-converter can be written
\begin{widetext}
\begin{equation} \label{eq:downconverter}
\ket{\psi_{out}} = \ket{0}_a \ket{0}_b + \int_{-\infty}^{\infty} \frac{2\chi\kappa}{\kappa ^2+\omega ^2}(\ket{0_\omega 1_{-\omega}}_a \ket{1_\omega 0_{-\omega}}_b + \ket{1_\omega 0_{-\omega}}_a \ket{0_\omega 1_{-\omega}}_b)\, \mathrm{d}\omega
\end{equation}
\end{widetext}
where $\chi$ is the conversion efficiency of the down-converter, and $\kappa$ is again related to the bandwidth of the output photons. We assume that conversion is very weak such that $\chi\ll\kappa$. This justifies dropping higher order terms involving higher photon numbers in Eq. (\ref{eq:downconverter}). The modes $a$ and $b$ correspond to the two spatial modes which form the outputs. It is evident from the form of Eq. (\ref{eq:downconverter}) that the down-conversion process does not always produce photon pairs, since it is possible that both outputs will be in the vacuum state. In fact, with the assumption that $\chi$ is small compared to $\kappa$, the down-converter will produce vacuum outputs most of the time. However, because the down-converter produces photon pairs if it produces anything at all, a reliable photon source can be constructed by conditioning on the detection of a photon at one output and using the other output as the source. Having performed conditioning, the normalized form of the photon produced at the other down-converter output is
\begin{equation} \label{eq:downconverter_distribution}
\alpha _\omega = \sqrt{\frac{\kappa}{2\pi\chi^2}} \frac{2\chi\kappa}{\kappa^2+\omega^2}
\end{equation}
where it has been assumed that the intrinsic response time of the counter, $\tau$, obeys $1/\tau\gg\kappa$ (see section \ref{sec:CNOT_gate}). Unlike the Lorentzian distribution, this distribution is a symmetric, real function in both the time and frequency domains. Passing a photon of this form through an appropriate filter can produce a Gaussian distributed packet. The normalized form of such a distribution is
\begin{equation} \label{eq:gaussian_distribution}
\alpha _\omega = \sqrt[4] {\frac{2}{\kappa^2\pi}} e^{-\frac{\omega^2}{\kappa^2}}
\end{equation}
where again $\kappa$ is related to the bandwidth of the distribution.

In addition to modeling arbitrary frequency distributions, we can also introduce an arbitrary time-delay onto a photon wave-packet. Using Fourier transform rules this can be modeled by introducing a complex exponential rotation factor into the frequency-domain distribution function. This corresponds to making the substitution $\alpha _\omega \to e^{i \omega \tau} \alpha _\omega$, where $\tau$ is the time-shift parameter. Thus, as an example, a Gaussian distribution with a time-shift would be expressed as
\begin{equation} \label{eq:time_shifted_gaussian}
\alpha _\omega = e^{i \omega \tau} \sqrt[4] {\frac{2}{\kappa^2\pi}} e^{-\frac{\omega^2}{\kappa^2}}
\end{equation}

Finally, mixed states can be represented by introducing a classical distribution of pure states. For example a single photon with inhomogeneous broadening of its frequency distribution can be represented by the density operator
\begin{widetext}
\begin{equation} \label{eq:mixed_state}
\hat{\rho} = \sqrt[4] {\frac{2}{\kappa^2\pi}} \int_{-\infty}^{\infty} d \Omega e^{-\frac{\Omega^2}{\kappa_{i}^2}} \int_{-\infty}^{\infty} \int_{-\infty}^{\infty} d \omega d \omega' \alpha_{\omega+\Omega} \alpha_{\omega'+\Omega}^{*} \ket{1}_{\omega}\bra{1}_{\omega'}
\end{equation}
\end{widetext}
In the following analysis we will restrict ourselves to pure input states.

\section{THE BEAMSPLITTER} \label{sec:beamsplitter}
\begin{figure}[!htb]
\includegraphics[width=0.45\textwidth]{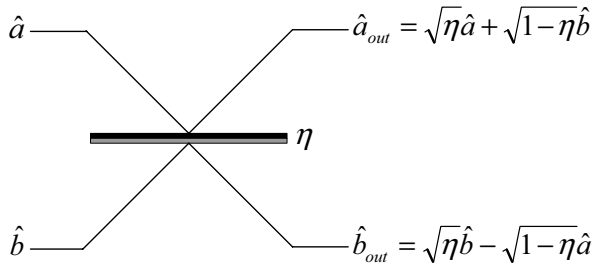}
\caption{\label{fig:beamsplitter}The phase asymmetric beamsplitter. Reflection off either surface or transmission from the `black' surface result in no sign change. Transmission from the `gray' surface results in a sign change.}
\end{figure}

To illustrate our method we begin by considering the operation of the beamsplitter, which is a fundamental building block in LOQC. We adopt the convention of the phase-asymmetric beamsplitter, shown in Fig. \ref{fig:beamsplitter}, which has the property that reflection off either surface or transmission from the `black' surface result in no sign change, whereas transmission from the `gray' surface results in a sign change.

The evolution of states through the beamsplitter is described in the Heisenberg picture by the equations of motion
\begin{eqnarray} \label{eq:beamsplitter}
\hat{\textbf{a}}_{out}=\sqrt{\eta}\hat{\textbf{a}}+\sqrt{1-\eta}\hat{\textbf{b}}\nonumber\\
\hat{\textbf{b}}_{out}=\sqrt{\eta}\hat{\textbf{b}}-\sqrt{1-\eta}\hat{\textbf{a}}
\end{eqnarray}
where $a$ and $b$ are the two spatial modes and $\eta$ is the beamsplitter's reflectivity.

\subsection{Calculating the beamsplitter output state}
We now describe how the Schr\"odinger picture evolution of the state may be obtained from Eq. (\ref{eq:beamsplitter}). First we express our input in terms of creation operators. For example, the input state with one photon at each beamsplitter input would be expressed as
\begin{equation} \label{eq:beamsplitter_input}
\ket{\psi}=\ket{1}_a\ket{1}_b \to \hat{\textbf{a}}^\dag\hat{\textbf{b}}^\dag\ket{0}_a\ket{0}_b
\end{equation}
To determine the corresponding output state for this input state we apply the time-evolution operator $\hat{U}$, to the input state. We also insert the identity operator $\hat{U}^\dag\hat{U}$, after each creation operator. Finally we recognize that when both inputs are in the vacuum state, both outputs must also be in the vacuum state. Hence we can apply the simplification $\hat{U}\ket{0}_a\ket{0}_b=\ket{0}_a\ket{0}_b$. Thus the output state can be represented as
\begin{eqnarray} \label{eq:beamsplitter_heisenberg}
\ket{\psi}_{out}&=&\hat{U}\ket{\psi}_{in}\nonumber\\
&=&\hat{U}\hat{\textbf{a}}^\dag\hat{\textbf{b}}^\dag\ket{0}_a\ket{0}_b\nonumber\\
&=&\hat{U}\hat{\textbf{a}}^\dag(\hat{U}^\dag\hat{U})\hat{\textbf{b}}^\dag(\hat{U}^\dag\hat{U})\ket{0}_a\ket{0}_b\nonumber\\
&=&(\hat{U}\hat{\textbf{a}}^\dag\hat{U}^\dag)(\hat{U}\hat{\textbf{b}}^\dag\hat{U}^\dag)\hat{U}\ket{0}_a\ket{0}_b\nonumber\\
&=&(\hat{U}\hat{\textbf{a}}^\dag\hat{U}^\dag)(\hat{U}\hat{\textbf{b}}^\dag\hat{U}^\dag)\ket{0}_a\ket{0}_b
\end{eqnarray}
The problem is now reduced to one of finding the reverse time evolved creation operators $\hat{U}\hat{\textbf{a}}^\dag\hat{U}^\dag$ and $\hat{U}\hat{\textbf{b}}^\dag\hat{U}^\dag$. We can determine these by first rewriting the equations of motion from Eq. (\ref{eq:beamsplitter}) as
\begin{eqnarray}
\hat{\textbf{a}}_{out}=\hat{U}^\dag\hat{\textbf{a}}\hat{U}=\sqrt{\eta}\hat{\textbf{a}}+\sqrt{1-\eta}\hat{\textbf{b}}\nonumber\\
\hat{\textbf{b}}_{out}=\hat{U}^\dag\hat{\textbf{b}}\hat{U}=\sqrt{\eta}\hat{\textbf{b}}-\sqrt{1-\eta}\hat{\textbf{a}}
\end{eqnarray}
By applying $\hat{U}$ from the left and $\hat{U}^\dag$ from the right of this expression we obtain
\begin{eqnarray}
\hat{\textbf{a}}=\sqrt{\eta}\hat{U}\hat{\textbf{a}}\hat{U}^\dag+\sqrt{1-\eta}\hat{U}\hat{\textbf{b}}\hat{U}^\dag\nonumber\\
\hat{\textbf{b}}=\sqrt{\eta}\hat{U}\hat{\textbf{b}}\hat{U}^\dag-\sqrt{1-\eta}\hat{U}\hat{\textbf{a}}\hat{U}^\dag
\end{eqnarray}
This results in a linear system of equations in terms of the reverse time evolved annihilation operators, which can be inverted to determine $\hat{U}\hat{\textbf{a}}\hat{U}^\dag$ and $\hat{U}\hat{\textbf{b}}\hat{U}^\dag$, and conjugated to determine $\hat{U}\hat{\textbf{a}}^\dag\hat{U}^\dag$ and $\hat{U}\hat{\textbf{b}}^\dag\hat{U}^\dag$. In this example the solution is
\begin{eqnarray}
\hat{U}\hat{\textbf{a}}^\dag\hat{U}^\dag=\sqrt{\eta}\hat{\textbf{a}}^\dag-\sqrt{1-\eta}\hat{\textbf{b}}^\dag\nonumber\\
\hat{U}\hat{\textbf{b}}^\dag\hat{U}^\dag=\sqrt{1-\eta}\hat{\textbf{a}}^\dag+\sqrt{\eta}\hat{\textbf{b}}^\dag
\end{eqnarray}
Substituting this result into the expression for the output state in Eq. (\ref{eq:beamsplitter_heisenberg}) and setting $\eta=0.5$ results in
\begin{eqnarray} \label{eq:beamsplitter_s}
\ket{\psi_{out}}&=&\Big(\frac{1}{2}\hat{\textbf{a}}^{\dag^2}-\frac{1}{2}\hat{\textbf{b}}^{\dag^2}\Big)\ket{0}_a\ket{0}_b\nonumber\\
&=&\frac{1}{\sqrt{2}}\ket{2}_a\ket{0}_b-\frac{1}{\sqrt{2}}\ket{0}_a\ket{2}_b
\end{eqnarray}
which displays quantum interference through complete suppression of the ``$\ket{1}_{a}\ket{1}_{b}$'' output term. This interference critically depends on the indistinguishability of the photons. We now generalize this calculation by allowing for differently distributed input states. We do this by making the substitution presented in Eq. (\ref{eq:distributed_state}). Thus the input state is now expressed as
\begin{eqnarray}
\ket{\psi}_{in}&=&\ket{1}_a\ket{1}_b\nonumber\\
&\to& \Big(\int \alpha _\omega \ket{1}_\omega \, \mathrm{d}\omega\Big)_a\Big(\int \beta_{\omega} \ket{1}_\omega \, \mathrm{d}{\omega}\Big)_b\nonumber\\
&=& \Big(\int \alpha _\omega \tilde{\textbf{a}}_\omega^\dag \, \mathrm{d}\omega \Big)\Big(\int \beta_{\omega}\tilde{\textbf{b}}_{\omega}^\dag \, \mathrm{d}{\omega}\Big)\ket{0}_a\ket{0}_b
\end{eqnarray}
It should be emphasized that this representation only allows for distinguishability in the time/frequency degree of freedom and does not allow for spectral entanglement. Upon applying the same procedure as before to this input state, we determine that the corresponding output state for the 50/50 beamsplitter is
\begin{eqnarray} \label{eq:non_ideal_beamsplitter_output}
\ket{\psi}_{out}&=&\ \ \frac{1}{2}\iint \alpha_\omega\beta_{\omega'}(\ket{1}_\omega\ket{1}_{\omega'})_a\ket{0}_b \, \mathrm{d}\omega\mathrm{d}\omega'\nonumber\\
&&-\frac{1}{2}\iint \alpha_\omega\beta_{\omega'}\ket{1}_{\omega,a}\ket{1}_{\omega',b} \, \mathrm{d}\omega\mathrm{d}\omega'\nonumber\\
&&+\frac{1}{2}\iint \alpha_{\omega'}\beta_\omega\ket{1}_{\omega,a}\ket{1}_{\omega',b} \, \mathrm{d}\omega\mathrm{d}\omega'\nonumber\\
&&-\frac{1}{2}\iint \alpha_\omega\beta_{\omega'}\ket{0}_a(\ket{1}_\omega\ket{1}_{\omega'})_b \, \mathrm{d}\omega\mathrm{d}\omega'
\end{eqnarray}
This result is completely general and allows us to substitute arbitrary frequency distribution functions. Without actually performing any substitutions however, we can gain much insight into the beamsplitter's behavior simply by inspecting the form of this result. The expression contains four terms, the first and last of which correspond to the terms in the ideal situation (see Eq. [\ref{eq:beamsplitter_s}]) where there are two photons present at one output and none at the other. The center two terms arise from photon distinguishability in which there is one photon present at each output. When $\alpha_\omega$ and $\beta_\omega$ are equal the center two terms cancel leaving only the ideal terms. This is the expected result, since so long as the inputs are indistinguishable ideal quantum behavior should ensue. In the opposing limit, when there is no overlap between $\alpha_\omega$ and $\beta_\omega$, no cancellation takes place and there will be a 50\% probability of measuring exactly one photon at each beamsplitter output. The photons then exhibit classical statistics.

From this expression for the output state we can calculate the coincidence rate $\langle\hat{N}_a\hat{N}_b\rangle$, between the outputs
\begin{eqnarray}
\langle\hat{N}_a\hat{N}_b\rangle&=&\frac{1}{2}\iint |\alpha_\omega|^2 |\beta_{\omega'}|^2 \, \mathrm{d}\omega\mathrm{d}\omega'\nonumber\\
&-&\frac{1}{2}\iint \alpha_\omega{\beta_\omega}^*{\alpha_{\omega'}}^*\beta_{\omega'} \, \mathrm{d}\omega\mathrm{d}\omega'
\end{eqnarray}
For $\alpha_\omega$ and $\beta_\omega$ normalized we obtain
\begin{equation} \label{eq:beamsplitter_expectation}
\langle\hat{N}_a\hat{N}_b\rangle=\frac{1}{2} - \frac{1}{2}\iint \alpha_\omega{\beta_\omega}^*{\alpha_{\omega'}}^*\beta_{\omega'} \, \mathrm{d}\omega\mathrm{d}\omega'
\end{equation}
Upon inspection we see that when $\alpha_\omega$ and $\beta_\omega$ are equal the two terms cancel, resulting in $\langle\hat{N}_a\hat{N}_b\rangle=0$ as expected for indistinguishable photons. On the other hand, when there is no overlap between $\alpha_\omega$ and $\beta_\omega$ the second term reduces to $0$, since this term calculates the overlap between the two distributions. Hence, assuming that the inputs are normalized, $\langle\hat{N}_a\hat{N}_b\rangle=0.5$ as expected for a classical distribution.

\subsection{Beamsplitter examples}
We now consider several examples of photons with different frequency distributions and arrival times.

\subsubsection{Gaussian distributions}
\begin{figure}[!htb]
\includegraphics[width=0.45\textwidth]{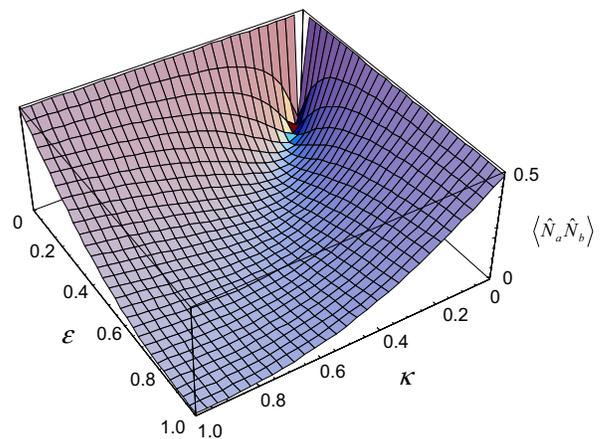}
\caption{\label{fig:NaNb_gaussian_freq}$\langle\hat{N}_a\hat{N}_b\rangle$ against the bandwidths of the two photons input to a beamsplitter, where both inputs are Gaussian distributed and arrive simultaneously. The units for $\varepsilon$ and $\kappa$ are arbitrary, provided they are consistent.}
\end{figure}

\begin{figure}[!htb]
\includegraphics[width=0.45\textwidth]{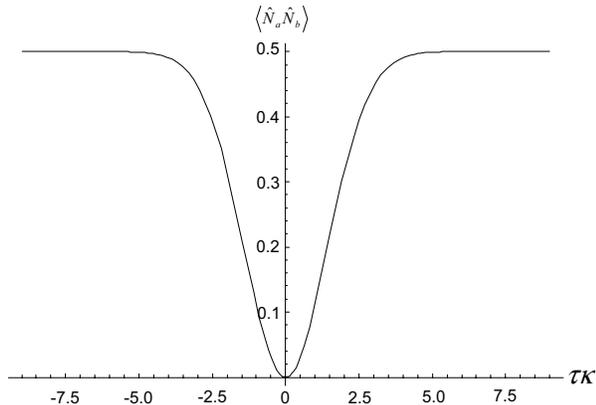}
\caption{\label{fig:NaNb_gaussian_time}$\langle\hat{N}_a\hat{N}_b\rangle$ against the time-shift introduced into one of the inputs to a beamsplitter, where both inputs are Gaussian distributed and share the same bandwidth $\kappa$.}
\end{figure}

Substituting Gaussian distributions (see Eq. [\ref{eq:gaussian_distribution}]) into the general expression for $\langle\hat{N}_a\hat{N}_b\rangle$ obtained in Eq. (\ref{eq:beamsplitter_expectation}) results in
\begin{equation} \label{eq:NaNb_gaussian}
\langle\hat{N}_a\hat{N}_b\rangle=\frac{1}{2}-\frac{\varepsilon\kappa}{\varepsilon^2+\kappa^2}
\end{equation}
where $\varepsilon$ and $\kappa$ are the bandwidth parameters associated with the two inputs. It is evident from this expression that when $\varepsilon=\kappa$, $\langle\hat{N}_a\hat{N}_b\rangle=0$. Alternately, if either $\varepsilon$ or $\kappa$ approach zero (but not both simultaneously) then $\langle\hat{N}_a\hat{N}_b\rangle=0.5$. This result is expected since a zero bandwidth distribution corresponds to a perfectly monochromatic photon, which will be completely distinguishable from any non-monochromatic photon. Fig. \ref{fig:NaNb_gaussian_freq} shows a plot of this behavior.
We now consider the case where a time-shift $\tau$, is introduced into one of the beamsplitter inputs. In this case we find that
\begin{equation} \label{eq:NaNb_gaussian_time}
\langle\hat{N}_a\hat{N}_b\rangle=\frac{1}{2}-\frac{1}{2}e^{-\frac{1}{4}\kappa^2\tau^2}
\end{equation}
where we have let $\varepsilon=\kappa$ for simplicity. Plotting this result yields Fig. \ref{fig:NaNb_gaussian_time}, which some readers will recognize as the familiar Hong-Ou-Mandel (HOM) \cite{bib:HOM87} dip. This result indicates that when no time-shift is present between identically distributed inputs, perfect quantum behavior arises. As the time-shift is increased and the two photons become increasingly distinguishable, the expectation value asymptotically approaches the classical result.

\subsubsection{Lorentzian distributions}
When using Lorentzian distributed inputs (see Eq. [\ref{eq:lorentzian_distribution}]) with different bandwidths, similar behavior arises to the Gaussian case. $\langle\hat{N}_a\hat{N}_b\rangle$ now has the form
\begin{equation}
\langle\hat{N}_a\hat{N}_b\rangle=\frac{1}{2}-\frac{2\varepsilon\kappa}{(\varepsilon+\kappa)^2}
\end{equation}
This result behaves almost identically to the Gaussian case and has the same properties in the limits.

\begin{figure}[!htb]
\includegraphics[width=0.45\textwidth]{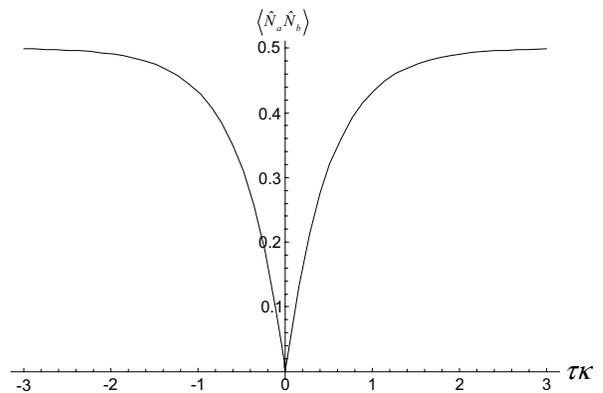}
\caption{\label{fig:NaNb_lorentzian_time}$\langle\hat{N}_a\hat{N}_b\rangle$ against the time-shift introduced into one of the inputs to a beamsplitter, where both inputs are Lorentzian distributed and share the same bandwidth $\kappa$.}
\end{figure}

Introducing a time-shift into one of the inputs results in
\begin{equation}
\langle\hat{N}_a\hat{N}_b\rangle=\frac{1}{2}-\frac{1}{2}e^{-2\kappa|\tau|}
\end{equation}
where we have again made the simplification $\varepsilon=\kappa$. Plotting this yields Fig. \ref{fig:NaNb_lorentzian_time}, which behaves similarly to the Gaussian case with the notable difference that the dip is no longer smooth, but rather sharp. This sharpness can be attributed to the discontinuity of the Lorentzian distribution in the time-domain.

\subsubsection{Down-conversion}
We now calculate $\langle\hat{N}_a\hat{N}_b\rangle$ when using inputs derived from conditioned down-converters. The result is of the form
\begin{equation}
\langle\hat{N}_a\hat{N}_b\rangle=\frac{1}{2}-\frac{2\varepsilon\kappa}{(\varepsilon+\kappa)^2}
\end{equation}
where we have let $\chi=1$ for simplicity. This result is identical to the Lorentzian case. Upon introducing a time-shift into one of the inputs we obtain
\begin{equation}
\langle\hat{N}_a\hat{N}_b\rangle=\frac{1}{2}-\frac{1}{2}e^{-2\kappa|\tau|}(1+\kappa|\tau|)^2 
\end{equation}
where $\varepsilon=\kappa$. The behavior of this result is indistinguishable from the Gaussian case, and, unlike the Lorentzian case, we again observe a smooth rather than a sharp dip.

\section{THE CONTROLLED-NOT GATE} \label{sec:CNOT_gate}
A number of non-deterministic, heralded two qubit gates have now been proposed \cite{bib:KLM01,bib:Knill02,bib:Pittman01}. We consider here the controlled-NOT (CNOT) gate proposed by Ralph \mbox{\emph{et.~al.}} \cite{bib:Ralph01}. This gate is simple in design and has previously been shown to exhibit good resilience to ancilla inefficiencies when compared to several other LOQC CNOT gate implementations \cite{bib:Lund03}.

The CNOT gate is the quantum equivalent of the classical XOR-gate. The gate operates on two input qubits, the control ($c$) and target ($t$), and implements the following transformations on the logical basis states $\ket{H}$ and $\ket{V}$
\begin{eqnarray} \label{eq:CNOT_transformations}
\ket{\psi}_{in}=\ket{H}_c\ket{H}_t \ &\to& \
\ket{\psi}_{out}=\ket{H}_c\ket{H}_t\nonumber\\
\ket{\psi}_{in}=\ket{H}_c\ket{V}_t \ &\to& \
\ket{\psi}_{out}=\ket{H}_c\ket{V}_t\nonumber\\
\ket{\psi}_{in}=\ket{V}_c\ket{H}_t \ &\to& \
\ket{\psi}_{out}=\ket{V}_c\ket{V}_t\nonumber\\
\ket{\psi}_{in}=\ket{V}_c\ket{V}_t \ &\to& \
\ket{\psi}_{out}=\ket{V}_c\ket{H}_t
\end{eqnarray}
where $\ket{H}\equiv\ket{1}_H\ket{0}_V$ and $\ket{V}\equiv\ket{0}_H\ket{1}_V$. Of course, the gate will also operate on superposition states.

\begin{figure*}[!tb]
\includegraphics[width=0.8\textwidth]{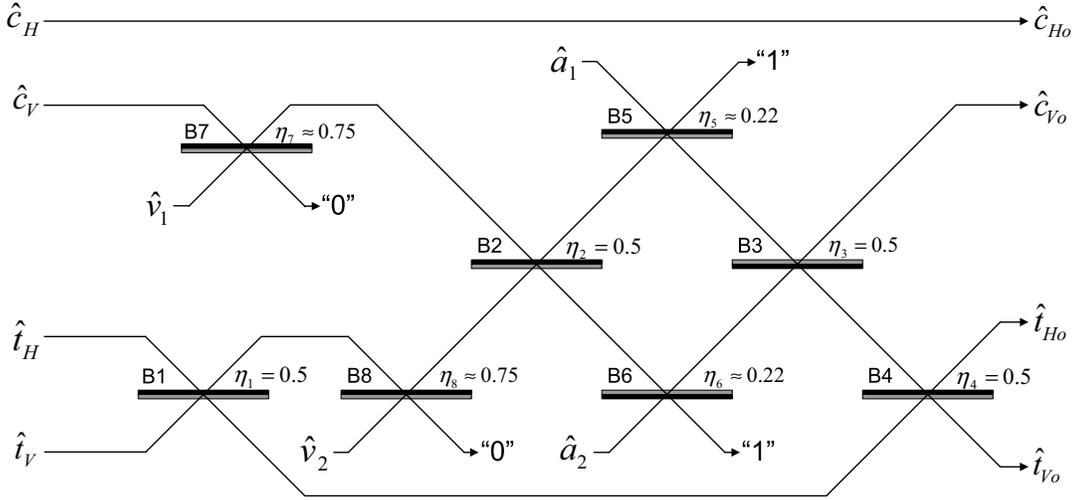}
\caption{\label{fig:CNOT_schematic}The layout of the simplified CNOT gate in the LOQC model.}
\end{figure*}

The gate, shown in Fig. \ref{fig:CNOT_schematic}, is implemented using eight beamsplitters, with reflectivities as indicated in the diagram ($\eta_1-\eta_8$). In addition to the control and target inputs the gate employs two vacuum ($v_1$, $v_2$) and two ancilla ($a_1$, $a_2$) inputs. The gate is non-deterministic, in that the success of the gate is conditional upon detecting no photon or exactly one photon at some of the outputs, labeled ``0'' and ``1'' respectively. Ideally the success probability is approximately 5\%.

\subsection{Modeling the CNOT gate}
In our study of the CNOT gate we will only consider photon distinguishability effects and ignore other possible experimental errors. Using the same approach as for the beamsplitter we derive a general expression for the input into the CNOT gate, where the inputs are subject to frequency distributions. The form of the input state is

\begin{eqnarray} \label{eq:CNOT_input_state}
\ket{\psi}_{in}&=&\Big[\alpha\Big(\int\kappa_\omega\tilde{\textbf{c}}_{H,\omega}^\dag\,\mathrm{d}\omega\Big)\Big(\int\sigma_\omega\tilde{\textbf{t}}_{H,\omega}^\dag\,\mathrm{d}\omega\Big)+\nonumber\\
&&\ \beta\Big(\int\kappa_\omega\tilde{\textbf{c}}_{H,\omega}^\dag\,\mathrm{d}\omega\Big)\Big(\int\sigma_\omega\tilde{\textbf{t}}_{V,\omega}^\dag\,\mathrm{d}\omega\Big)+\nonumber\\
&&\ \gamma\Big(\int\kappa_\omega\tilde{\textbf{c}}_{V,\omega}^\dag\,\mathrm{d}\omega\Big)\Big(\int\sigma_\omega\tilde{\textbf{t}}_{H,\omega}^\dag\,\mathrm{d}\omega\Big)+\nonumber\\
&&\ \delta\Big(\int\kappa_\omega\tilde{\textbf{c}}_{V,\omega}^\dag\,\mathrm{d}\omega\Big)\Big(\int\sigma_\omega\tilde{\textbf{t}}_{V,\omega}^\dag\,\mathrm{d}\omega\Big)\Big]\nonumber\\
&&\ \Big(\int\varepsilon_\omega\tilde{\textbf{a}}_{1,\omega}^\dag\,\mathrm{d}\omega\Big)\Big(\int\varepsilon_\omega\tilde{\textbf{a}}_{2,\omega}^\dag\,\mathrm{d}\omega\Big)\ket{\textbf{0}}
\end{eqnarray}

where $\alpha$, $\beta$, $\gamma$ and $\delta$ determine the input superposition, and $\kappa_\omega$, $\sigma_\omega$ and $\varepsilon_\omega$ are the frequency distribution functions of the control, target and ancilla inputs respectively.

Using this expression for a completely general input state, we wish to determine the corresponding output state. We use a similar approach as previously for the beamsplitter. Firstly, we derive the equations of motion for the system by recursively using the equations of motion for a single beamsplitter. This system of equations is then inverted to determine the reverse time evolved creation operators for each of the input modes. These operators are used to evolve the input state given in Eq. (\ref{eq:CNOT_input_state}). We omit the resulting expression due to its complexity.

\subsection{Modeling photon counters}
Unlike the beamsplitter we considered previously, the CNOT gate involves a conditioning process which must also be modeled. This is performed using photon counters. In a standard model, in which frequency distributions of inputs are ignored, the conditioning process could be modeled by applying a series of projectors representing the desired measurements. In the case of the CNOT gate this would be expressed as
\begin{equation} \label{eq:ideal_CNOT_projector}
(\ket{0}_{v_1}\bra{0}_{v_1}) \otimes (\ket{0}_{v_2}\bra{0}_{v_2}) \otimes (\ket{1}_{a_1}\bra{1}_{a_1}) \otimes (\ket{1}_{a_2}\bra{1}_{a_2})
\end{equation}
This sequence of projectors effectively discards terms which contain photons at any of the vacuum outputs or not exactly one photon at either of the ancilla outputs, the desired post-selection procedure. Clearly this sequence of projectors is incompatible with our frequency-distributed model and must be modified.

In the frequency distribution picture we modify the projector from Eq. (\ref{eq:ideal_CNOT_projector}) to
\begin{eqnarray} \label{eq:CNOT_projector}
&&(\ket{0}_{v_1}\bra{0}_{v_1}) \otimes (\ket{0}_{v_2}\bra{0}_{v_2}) \otimes \Big(\int\ket{1}_{a_1,\omega}\bra{1}_{a_1,\omega}\,\mathrm{d}\omega\Big) \otimes \nonumber\\
&&\Big(\int\ket{1}_{a_2,\omega}\bra{1}_{a_2,\omega}\,\mathrm{d}\omega\Big)
\end{eqnarray}
This projector describes post-selection using ideal photon counters. We now generalize the model further to accommodate for two important real-life effects, the counters' frequency and time resolution.

Any photon counter will invariably be sensitive to only a finite range of frequencies. More generally we can introduce a function, say $\mu_\omega$, which describes this frequency response. We model the counter's frequency selectivity with the introduction of a filter prior to the projector. The filter is described by the equations of motion
\begin{eqnarray}
\tilde{\textbf{a}}_{out,\omega}=\sqrt{\mu_\omega}\tilde{\textbf{a}}_\omega+\sqrt{1-\mu_\omega}\tilde{\textbf{b}}_\omega\nonumber\\ 
\tilde{\textbf{b}}_{out,\omega}=\sqrt{\mu_\omega}\tilde{\textbf{b}}_\omega-\sqrt{1-\mu_\omega}\tilde{\textbf{a}}_\omega 
\end{eqnarray}
In this model the counter's input enters through $a$ while $b$ is kept in the vacuum state. $a_{out}$ then contains the transmitted (or accepted), and $b_{out}$ the reflected component of the input state. The function $\mu_\omega$ varies between 0 and 1, where 0 indicates no sensitivity, and 1 complete sensitivity, to the respective frequency component. We would expect that this function would generally have a Gaussian-like distribution. The components which are reflected out of the $b$-mode are `unseen' by the observer. Hence this filtering process results in mixing effects. Therefore, following this filtering stage the state of the system must be expressed in density operator form, where the reflected modes are traced out to produce a reduced density operator describing the observed system. Thus
\begin{equation}
\hat{\rho}_{filtered}=\mathrm{tr}_B(\ket{\psi_{filtered}}\bra{\psi_{filtered}})
\end{equation}
The integral form projector (Eq. [\ref{eq:CNOT_projector}]) is then applied to this reduced density operator.

In addition to finite frequency selectivity, in any experiment the photon counters will only be `open' for a finite time period. We model this time selectivity in an analogous manner to our modeling of frequency selectivity, except that the equations of motion describing the filtering process are now in the time-domain rather than the frequency-domain
\begin{eqnarray}
\hat{\textbf{a}}_{out,\tau}=\sqrt{\eta_\tau}\hat{\textbf{a}}_\tau+\sqrt{1-\eta_\tau}\hat{\textbf{b}}_\tau\nonumber\\
\hat{\textbf{b}}_{out,\tau}=\sqrt{\eta_\tau}\hat{\textbf{b}}_\tau-\sqrt{1-\eta_\tau}\hat{\textbf{a}}_\tau
\end{eqnarray}
where $\eta_\tau$ is a time dependent function, varying between 0 and 1, which describes the time-response of the counter. In general we
would expect this function to be approximately rectangular, mimicking the opening and closing of the shutter.

\subsection{Characterizing CNOT gate behavior}
Having developed an expression for the output state of the CNOT gate for a general input state and photon counter response we wish to characterize the behavior of the gate as various parameters are changed. We use two quantifiers to do this, the fidelity and success probability of the gate.

The fidelity is a measure of how close the actual output state and expected output state of the gate are, where the expected output state is defined according to the logical transformations given in Eq. (\ref{eq:CNOT_transformations}), and we assume that all inputs share identical distributions and arrive simultaneously. Hence, the fidelity quantifies how accurately the gate is performing the desired logical transformation. Mathematically this is defined as
\begin{equation}
F=\frac{\bra{\psi_{exp}}\hat{\rho}_{out}\ket{\psi_{exp}}}{\mathrm{tr}(\hat{\rho}_{out})}
\end{equation}
where $\ket{\psi}_{exp}$ is the normalized expected output state and $\hat{\rho}_{out}$ the density operator of the actual output state. The trace is introduced into the denominator to normalize $\hat{\rho}_{out}$, thereby decoupling $F$ from the success probability. Due to normalization we expect $F$ to vary between 0 and 1, where 1 corresponds to ideal gate behavior, and 0 to completely non-ideal behavior. We expect to see $F=1$ when all of the input photons are indistinguishable and the counters are behaving ideally (\mbox{\emph{i.e.}} they have infinite bandwidths and time windows).

We should caution that the use of a 1-dimensional measure such as fidelity to characterize our gate rolls a number of effects into a single number. In particular $\ket{\psi_{exp}}$ is defined as the expected output state when the input states have identical distributions and arrive simultaneously. This has the effect that if, for example, a time-shift were introduced into one of the inputs then $F$ would suffer as a result of two separate effects. The obvious effect is that the gate would no longer be operating ideally and thus the transformation implemented by the gate would be different to what is expected. The less obvious effect is that even if the logical transformation implemented by the gate were not changed, $F$ would still suffer since $\ket{\psi_{exp}}$ is defined as the output state when no time-shifts are present. Hence we are calculating the overlap of a state which is displaced temporally with one which is not. There is no obvious way to decouple the measurement of these two effects. On the other hand we should remember that ultimately multiple quantum gates are intended to be integrated into quantum circuits. In the context of a circuit whereby the output from our gate is feeding into the input of another, both of the aforementioned effects are degrading and it is therefore valid that $F$ takes both into account.

The success probability measures how often the conditioning process, and hence gate operation, succeeds. This is defined as
\begin{equation}
P=\mathrm{tr}(\hat{\rho}_{out})
\end{equation}
Thus the success probability is simply a reflection of the normalization factor of the output state. Recall that in the ideal case the CNOT gate succeeds approximately 5\% of the time. Hence in the ideal limit we expect $P$ to approach 0.05.

Using these two measures we can effectively characterize how the behavior of the CNOT gate varies for any given input superposition as the input distribution functions and counter parameters are varied. Gate performance is input state dependent, so instead of considering a particular superposition we consider the worst-case scenario whereby we search across all possible input superpositions and determine the minimum values for both the fidelity and success probability ($F_{min}$ and $P_{min}$). Hence
\begin{eqnarray}
F_{min}=\min(F)\ \ \forall \ \ \alpha,\beta,\gamma,\delta \nonumber\\
P_{min}=\min(P)\ \ \forall \ \ \alpha,\beta,\gamma,\delta
\end{eqnarray}
In our studies we perform the searches using a Monte Carlo approach.

\subsection{CNOT gate with time-shifted Gaussian inputs}
From the beamsplitter examples in section \ref{sec:beamsplitter} we observe that qualitatively the results are similar across the different input distributions considered. For simplicity we now restrict ourselves to considering Gaussian distributed inputs and postulate that different distributions will yield similar results.

\begin{figure*}[!tb]
\includegraphics[width=1.0\textwidth]{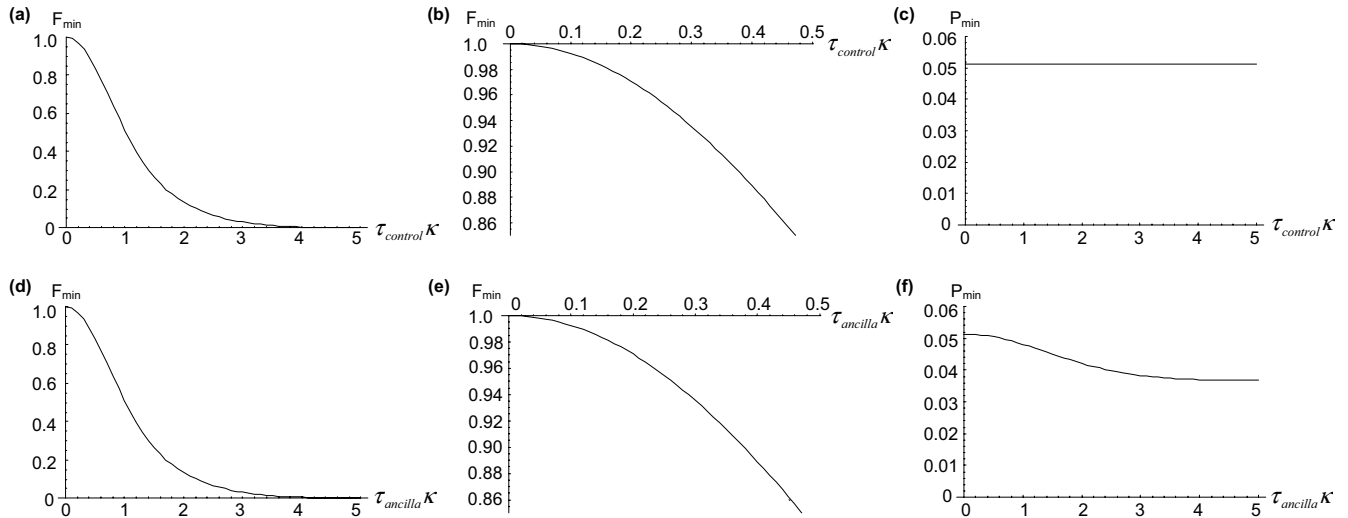}
\caption{\label{fig:CNOT_results}Plots of $F_{min}$ (a,b,d,e) and $P_{min}$ (c,f) for the CNOT gate as the time-shift parameters of the control (top) and ancilla (bottom) inputs are varied. Plots (b) and (e) are fidelity plots, zoomed in on the region of high fidelity ($F>0.85$). All inputs are Gaussian distributed with bandwidth $\kappa$.}
\end{figure*}

Time-shifted Gaussian distributions (see Eq. [\ref{eq:time_shifted_gaussian}]) were substituted for all of the frequency distribution functions in the expression for the CNOT gate's input state, with all bandwidth parameters equal. Hence, the inputs into the gate differed only in their arrival time. We substituted flat distributions for the photon counters' frequency and time response functions. Thus the counters were ideal. Using these parameters, $F_{min}$ and $P_{min}$ were calculated as each of the time-shift parameters were varied. Fig. \ref{fig:CNOT_results} illustrates the results when the $\tau_{control}$ (top) and $\tau_{ancilla}$ (bottom) parameters are varied. It should be noted that, in the context of a search across all possible input states, varying $\tau_{target}$ is equivalent to varying $\tau_{control}$, since beamsplitters B1 and B4 are invisible to such a search and thus the gate becomes completely symmetric. Similarly, varying $\tau_{control}$ and $\tau_{target}$ in unison is equivalent to varying $\tau_{ancilla}$.

It is immediately obvious that the fidelity plots behave as expected in the limits. Specifically, when no time-shift is present in the input in question, perfect fidelity is observed. As the time-shift in the input approaches infinity, the fidelity asymptotically approaches zero (\mbox{\emph{i.e.}} the output of the gate no longer bears any resemblance to what is expected). Similarly, when no time-shift is present we observe the ideal-case success probability of approximately 5\%, which decreases as the time-shift is increased. Note however, that in general the success probability does not asymptotically approach zero. In fact, when either the control or target inputs are displaced (but not both simultaneously) $P_{min}$ is completely unaffected. If $P$ is plotted for specific superpositions we often see that $P$ actually increases as time-shift parameters are increased. Current estimates place the fault-tolerant threshold for LOQC at the 1\% level \cite{bib:Knill04}. Our results imply that time synchronization needs to be to better than $1/10$ of the inverse frequency bandwidth ($1/\kappa$) to satisfy this threshold. Equivalently we can estimate that inhomogeneous temporal broadening (time jitter) must have a standard deviation less than this figure to reach fault tolerance. This result is somewhat better in terms of tolerable jitter than that obtained by Kiraz \mbox{\emph{et.~al.}} \cite{bib:Kiraz04} for Lorentzian inputs to the Knill gate \cite{bib:Knill02}.

Although we have performed our calculations explicitly as time-displaced inputs, transform symmetry means completely equivalent graphs would be generated as a function of frequency displacements and so we can similarly estimate that inhomogeneous frequency broadening should be less than $1/10$ of the inverse temporal bandwidth. More generally any variation in the distinguishability of the input qubits can be mapped to the results presented here. This is because it is inherently the degree of distinguishability between input states which determines the effectiveness of the gate's operation, not the specific nature or source of the distinguishability. The conclusion is that although the results presented are very specific, their interpretation is very general and the parameters $\tau_{control}$, $\tau_{target}$ and $\tau_{ancilla}$ can be more broadly interpreted as `distinguishability factors' than specifically as time-shifts. We note that although the results presented are for pure inputs states, results for mixed input states (see Eq. [\ref{eq:mixed_state}]) could be obtained by performing a weighted average over the results shown.

\subsection{CNOT gate with non-ideal photon counters}
\begin{figure*}[!htb]
\includegraphics[width=0.7\textwidth]{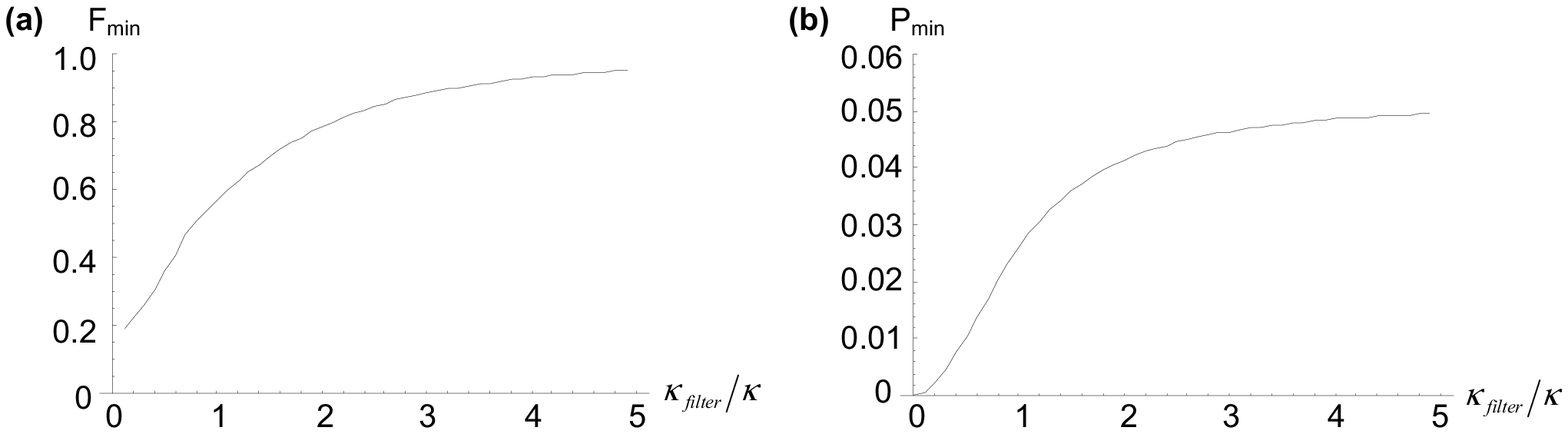}
\caption{\label{fig:CNOT_detector_bandwidth}$F_{min}$ (a) and $P_{min}$ (b) against the bandwidth of the photon counters. The counters operate over an infinite time-range. All inputs are Gaussian distributed with bandwidth $\kappa$ and simultaneous arrival time.}
\end{figure*}

We finally consider the effect upon the operation of the CNOT gate of utilizing photon counters which have finite frequency or time resolution. Initially we examine frequency resolution effects by substituting a Gaussian distribution for the frequency response function $\mu_\omega$, and leaving the time-range over which the counters operate as infinite. We assume that all of the input states are indistinguishable (\mbox{\emph{i.e.}} $\tau_{control}=\tau_{target}=\tau_{ancilla}=0$). Fig. \ref{fig:CNOT_detector_bandwidth} shows a plot of the worst-case fidelity (a) and success probability (b) against photon counter bandwidth.
As the counter bandwidth approaches infinity, both the fidelity and success probability approach the ideal limit. In the opposing limit, as the counters becomes completely selective, the success probability approaches zero and the fidelity approximately 0.2. In other words, the gate never succeeds and the gate's output is essentially random.

\begin{figure*}[!htb]
\includegraphics[width=0.7\textwidth]{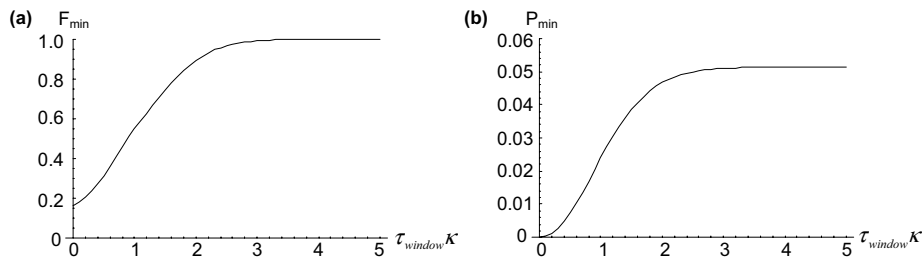}
\caption{\label{fig:CNOT_detector_window}$F_{min}$ (a) and $P_{min}$ (b) against the size of the time-window of the photon counters. The counters have infinite bandwidth. All inputs are Gaussian distributed with bandwidth $\kappa$ and simultaneous arrival time.}
\end{figure*}

Next we examine time resolution effects by assuming that the photon counters have infinite bandwidth, but operate only over a finite time period. We model this by substituting a rectangular function, ranging from $-\tau_{window}$ to $\tau_{window}$, for the time response function $\eta_\tau$. This is intended to model the behavior of a counter with a shutter with very fast reaction time. Hence the transient behavior of the shutter is ignored. The results are shown in Fig. \ref{fig:CNOT_detector_window}, which exhibits similar behavior to the plots against the counters' bandwidths.

\begin{figure}[!htb]
\includegraphics[width=0.35\textwidth]{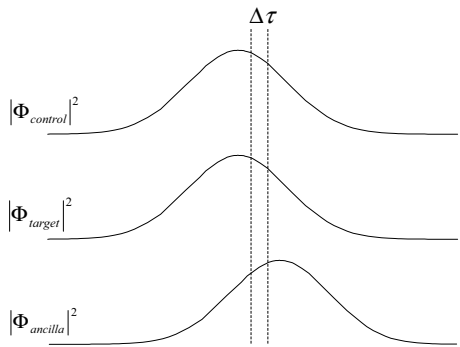}
\caption{\label{fig:window_tuning}A plot of the arriving photon wave-packets, where there is a time displacement of the ancilla photons relative to the other two photons. The region $\Delta\tau$ is where wave-function overlap is maximal and the photons are almost indistinguishable.}
\end{figure}

Legero \mbox{\emph{et.~al.}} \cite{bib:Legero03} recently performed a HOM-like experiment in which photons were intentionally introduced with a relative time-shift. It was found that if the time resolution of the photon counters was sufficiently narrow compared to the photons' length, photon distinguishability could be compensated for by `zooming in' on the time period where the overlap between the photons' wave-functions was maximal and the photons were effectively indistinguishable. One might question whether a similar approach is possible in the context of the CNOT gate to improve fidelity at the expense of success probability. Fig. \ref{fig:window_tuning} illustrates this concept.
Upon investigation it is evident that in general this approach does not work for the implementation of the CNOT gate described. This is due to photon-number ambiguity at the counters. In the HOM experiment it is known, due to conservation of the number of photons in the system, that if both counters `click' exactly once then exactly one photon must have been incident on each counter. In the context of the CNOT gate we cannot make such assumptions due to the extra ancillary photons which make it possible that anywhere between zero and three photons will be incident upon the counters. Hence, if a counter with a finite time window clicks once then this does not necessarily imply that exactly one photon was incident upon the counter. It is possible that two or even three photons were incident, but were simply not seen due to the finite time window. For this reason we postulate that in general it is not possible to improve the operation (\mbox{\emph{i.e.}} fidelity) of the CNOT gate by manipulating the time window of the conditioning photon counters. We note that this argument does not apply to current demonstrations of the CNOT gate which operate in coincidence, as photon-number ambiguity is no longer an issue. Calculations on the O'Brien \mbox{\emph{et.~al.}} gate \cite{bib:Ralph02,bib:OBrien03} show that fidelity can be improved arbitrarily close to unity, at the expense of success probability, for arbitrary temporal mismatch between the inputs. The reasoning applied here to temporal filtering, when applied in the frequency domain, leads to the well known advantages of frequency filtering in experiments which operate in coincidence.

\section{\label{sec:conclusion}CONCLUSION}
In this paper we have developed a general formalism for including frequency and temporal effects into the evaluation of LOQC networks. To illustrate our approach we looked at how non-classical interference at a beamsplitter is affected by differing frequency distributions and time shifts. As an example of an LOQC circuit we examined the effect of time-shifts on a non-deterministic, heralded CNOT gate. We concluded that fidelities higher than 99\% required photon synchronization to better than 10\% of the inverse bandwidth. Similar results will apply to frequency shifts. Finally we examined the effects of non-ideal photon counters upon the operation of the CNOT gate, specifically considering the time and frequency response of the counters. We found that in general the fidelity of the gate diminishes as either the frequency or time bandwidth of the counters decreases. Furthermore, the gate's fidelity cannot be improved through manipulation of the photon counters' frequency or time-responses. This is in contrast to the situation with gates which operate in coincidence where such manipulations can be quite effective.

\begin{acknowledgments}
We wish to thank Gerard Milburn and Tamyka Bell for fruitful discussions. This work was supported by the Australian Research Council.
\end{acknowledgments}

\bibliography{paper.bib}

\end{document}